\documentclass[useAMS,usenatbib]{mn2e}
\usepackage{lmacs}
\usepackage{graphicx}
\usepackage[T1]{fontenc} 
\usepackage{aecompl}
\usepackage{aas_macros}
\usepackage{color, soul}
\usepackage{gensymb}
\usepackage{amsmath}
\usepackage{url}
\graphicspath{{./figs/}}
\def\altaffilmark#1{$^{#1}$}
\def\altaffiltext#1#2{$^{#1}$#2}
\newcounter{aaffilcoun}

\newcounter{affilcoun}

\usepackage{xargs}

\newcommand{\name}{ASASSN-18tb\xspace}

\newcommand{\Ha}{H\ensuremath{\alpha}\xspace}

\newcommand{\sne}{SNe Ia\xspace}

\title[ASASSN-18tb]{ASASSN-18tb: A Most Unusual Type Ia Supernova Observed by \textit{TESS} and SALT}

\author[Vallely et al.]
{P. J. Vallely\altaffilmark{1},
M. Fausnaugh\altaffilmark{2,3},
S. W. Jha\altaffilmark{4,5},
M. A. Tucker\altaffilmark{6},
Y. Eweis\altaffilmark{4},
B. J. Shappee\altaffilmark{6},
\newauthor
C. S. Kochanek\altaffilmark{1,7},
K. Z. Stanek\altaffilmark{1,7},
Ping Chen\altaffilmark{8},
Subo Dong\altaffilmark{8},
J. L. Prieto\altaffilmark{9,10},
\newauthor
T. Sukhbold\altaffilmark{1,7},
Todd A. Thompson\altaffilmark{1,7,11},
J. Brimacombe\altaffilmark{12},
M. D. Stritzinger\altaffilmark{13},
\newauthor
T. W.-S. Holoien\altaffilmark{14},
D. A. H. Buckley\altaffilmark{15},
M. Gromadzki\altaffilmark{16},
Subhash Bose\altaffilmark{8}
\\
\altaffiltext{1}{Department of Astronomy, The Ohio State University, 140 West 18th Avenue, Columbus, OH 43210, USA} \\
\altaffiltext{2}{Department of Physics, Massachusetts Institute of Technology, Cambridge, MA 02139, USA} \\
\altaffiltext{3}{Kavli Institute for Astrophysics and Space Research, Massachusetts Institute of Technology, Cambridge, MA 02139, USA} \\
\altaffiltext{4}{Department of Physics and Astronomy, Rutgers, the State University of New Jersey,136 Frelinghuysen Rd., Piscataway, NJ 08854, USA} \\
\altaffiltext{5}{Center for Computational Astrophysics, Flatiron Institute, 162 5th Avenue, New York, NY 10010, USA} \\
\altaffiltext{6}{Institute for Astronomy, University of Hawai'i, 2680 Woodlawn Drive, Honolulu, HI 96822, USA} \\
\altaffiltext{7}{Center for Cosmology and AstroParticle Physics, The Ohio State University, 191 W. Woodruff Ave., Columbus, OH 43210, USA} \\
\altaffiltext{8}{Kavli Institute for Astronomy and Astrophysics, Peking University, Yi He Yuan Road 5, Hai Dian District, Beijing 100871, China}\\
\altaffiltext{9}{N\'ucleo de Astronom\'ia de la Facultad de Ingenier\'ia, Universidad Diego Portales, Av. Ej\'ercito 441, Santiago, Chile} \\
\altaffiltext{10}{Millennium Institute of Astrophysics, Santiago, Chile} \\
\altaffiltext{11}{Institute for Advanced Study, 1 Einstein Drive, Princeton, NJ 08540, USA} \\
\altaffiltext{12}{Coral Towers Observatory, Cairns, Queensland 4870, Australia} \\
\altaffiltext{13}{Department of Physics and Astronomy, Aarhus University, Ny Munkegade 120, DK-8000 Aarhus C, Denmark} \\
\altaffiltext{14}{Carnegie Observatories, 813 Santa Barbara Street, Pasadena, CA 91101, USA}\\
\altaffiltext{15}{South African Astronomical Observatory, PO Box 9, Observatory 7935, Cape Town, South Africa}\\
\altaffiltext{16}{Warsaw University Astronomical Observatory, Al. Ujazdowskie 4, 00-478 Warszawa, Poland}
\\}

\begin{document}

\date{Accepted xxx Received xx; in original form xxx}

\pagerange{\pageref{firstpage}--\pageref{lastpage}} \pubyear{2019}

\maketitle

\label{firstpage}

\begin{abstract}

We present photometric and spectroscopic observations of the unusual Type Ia supernova ASASSN-18tb, including a series of SALT spectra obtained over the course of nearly six months and the first observations of a supernova by the \textit{Transiting Exoplanet Survey Satellite (TESS)}. We confirm a previous observation by \cite{2019Kollmeier} showing that ASASSN-18tb is the first relatively normal Type Ia supernova to exhibit clear broad ($\sim1000$ km s$^{-1}$) H$\alpha$ emission in its nebular phase spectra. We find that this event is best explained as a sub-Chandrasekhar mass explosion producing $M_{Ni} \approx 0.3\; \rm{M}_\odot$. Despite the strong H$\alpha$ signature at late times, we find that the early rise of the supernova shows no evidence for deviations from a single-component power-law and is best fit with a moderately shallow power-law of index $1.69\pm0.04$. We find that the H$\alpha$ luminosity remains approximately constant after its initial detection at phase +37 d, and that the H$\alpha$ velocity evolution does not trace that of the Fe~III$~\lambda4660$ emission. These suggest that the H$\alpha$ emission arises from a circumstellar medium (CSM) rather than swept up material from a non-degenerate companion. However, ASASSN-18tb is strikingly different from other known CSM-interacting Type Ia supernovae in a number of significant ways. Those objects typically show an H$\alpha$ luminosity two orders of magnitude higher than what is seen in ASASSN-18tb, pushing them away from the empirical light curve relations that define ``normal'' Type Ia supernovae. Conversely, ASASSN-18tb exhibits a fairly typical light curve and luminosity for an underluminous or transitional SN~Ia, with $M_R \approx -18.1$ mag. Moreover, ASASSN-18tb is the only SN~Ia showing H$\alpha$ from CSM interaction to be discovered in an early-type galaxy.
\end{abstract}

\section{Introduction}

\begin{figure*}
\includegraphics[width=0.49\textwidth]{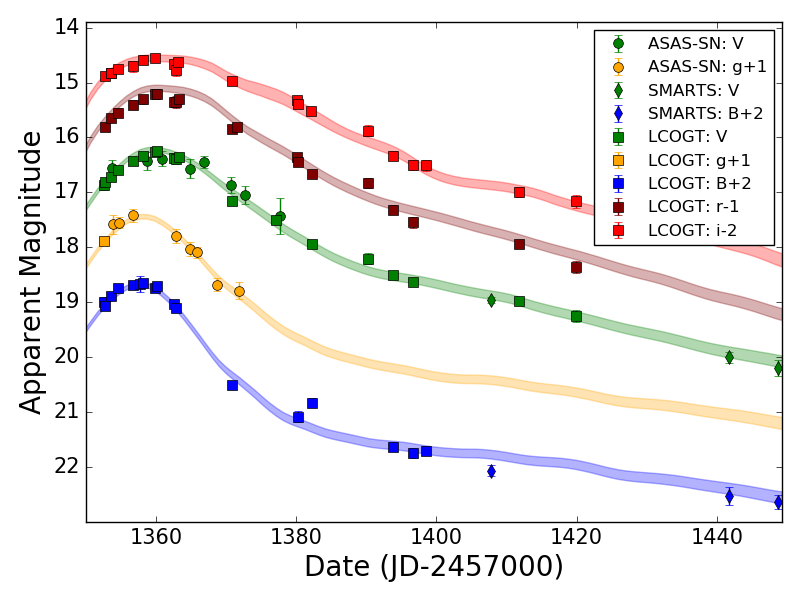}
\hfill
\includegraphics[width=0.49\textwidth]{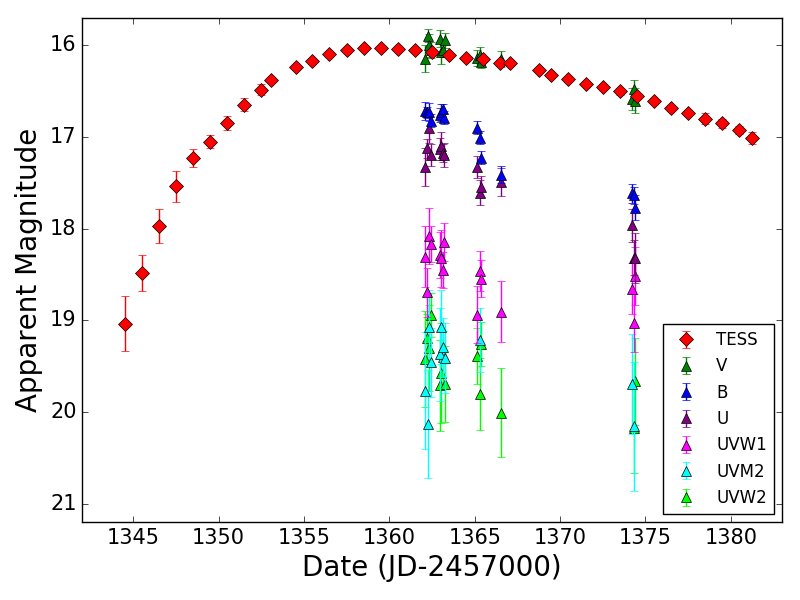}
\caption{Photometric observations of ASASSN-18tb. Ground-based $BVgri$ photometry obtained with LCOGT, ASAS-SN, and  SMARTS is shown on the left, and space-based \textit{TESS} and \textit{Swift} UVOT photometry is shown on the right. The \textit{TESS} photometry is shown for 24 hour bins. Marker colors indicate the filter, and marker shapes indicate the source of the data. Error bars are shown for all points, but can be smaller than the symbol used to represent the data. The photometry is not corrected for Galactic extinction. The shaded bands in the left panel show the MLCS2k2 fit \citep{2007Jha} to the LCOGT light curve.}
\label{fig:GroundandSpaceLightCurve}
\end{figure*}

It has been known for some time that Type Ia supernovae (SNe~Ia) are the result of the thermonuclear explosion of a carbon-oxygen white dwarf (CO WD) triggered by a companion \citep{1960Hoyle,1969Colgate,2011Nugent}. However, the physical nature of this companion and the details of the explosion mechanism remain an active area of debate. Broadly speaking, SN~Ia progenitor models can be grouped into two categories -- the single-degenerate (SD) and double-degenerate (DD) scenarios.

In the standard DD scenario, a tight WD-WD binary loses energy and angular momentum via gravitational wave emission before undergoing tidal interactions and subsequently exploding  \citep{1979Tutukov,1984Iben,1984Webbink,2012Shen}.
\cite{2011Thompson} proposed that SNe~Ia originate from triple systems, and showed that Lidov-Kozai oscillations driven by a tertiary companion can accelerate WD-WD mergers via gravitational wave radiation and implied that they may lead to WD-WD collisions. \cite{2012Katz} and \cite{2013Kushnir} proposed and found supporting evidence suggesting that WD-WD direct collisions in triple systems may be a major channel for SNe Ia.
Further evidence for SNe~Ia produced through this scenario has been found by \cite{2015Dong} and \cite{2019Vallely} in the form of bimodal distributions of $^{56}$Ni decay products in nebular phase spectra.

In the canonical SD scenario, the WD accretes matter from a non-degenerate stellar companion, eventually approaching the Chandrasekhar limit and undergoing a thermonuclear runaway \citep{1973Whelan,1982Nomoto,2004Han}. The stellar companion will be struck by the supernova ejecta shortly after explosion, leading directly to a number of observable signatures. First, the companion interaction should lead to excess emission in the early-phase light curve. Although this emission is strongly dependent on the characteristics of the stellar companion and the viewing angle of the system, \cite{2010Kasen} showed that it should be observable for an appreciable number of SNe~Ia. Additionally, material stripped from the companion star should produce hydrogen emission lines visible in late-time nebular spectra \citep{1975Wheeler,2000Marietta,2012Pan,2012Liu,2017Boehner}. Finally, the ejecta interactions impact the post-explosion properties of the stellar companion (see, e.g., \citealt{2003Podsiadlowski}, \citealt{2012bPan}, and \citealt{2013bShappee}).

Early time observations are being obtained for steadily increasing numbers of SNe~Ia.  Most of these efforts have focused on finding or placing upper limits on excess emission due to ejecta colliding with a nearby SD companion, although \citet{2018Stritzinger} have also found evidence that the early time optical colors are correlated with the post-peak decline rates. The searches for distortions in the early time light curves have produced mixed results. Many SN~Ia light curves do not show evidence of companion interaction. The nearby Type Ia SN~2011fe had an early-phase light curve consistent with a single-component power-law \citep{2011Nugent, 2012Bloom}, and early-time observations of SN~2009ig are inconsistent with the \cite{2010Kasen} interaction models \citep{2012Foley}. Additionally, \cite{2015Olling} found no evidence for ejecta-companion interaction when examining three SNe~Ia observed by \textit{Kepler} \citep{2010Borucki}. Based on early excess non-detections, \cite{2016Shappee} were able to rule out most non-degenerate companions for ASASSN-14lp, and \cite{2018Holmbo} were able to place even tighter constraints on SN~2013gy.

However, this is not the case for all events. An early linear phase in the light curve of SN~2013dy was observed by \cite{2013Zheng}, and observations of SN~2014J show evidence for additional early-time structure \citep{2014Zheng,2015Goobar,2015Siverd}. \cite{2018Contreras} found that the light curve of SN~2012fr had an initial roughly linear phase that lasted for $\sim2.5$ days, and \textit{K2} observations of ASASSN-18bt showed a similar $\sim4$ day linear phase \citep{2018Brown,2019Shappee,2019Li,2019Dimitriadis}. Additionally, \cite{2016Marion} found potential indications of interaction with a non-degenerate binary companion in SN~2012cg, although this interpretation is challenged by \cite{2018Shappee}.

Searches for hydrogen emission lines at late times as evidence for stripped material have largely failed. No such signatures were detected for SNe 1998bu and 2000cx \citep{2013Lundqvist}, SN~2001el \citep{2005Mattila}, SNe 2005am and 2005cf \citep{2007Leonard}, SN~2012cg \citep{2018Shappee}, SN~2013gy \citep{2018Holmbo}, or SN~2017cbv \citep{2018Sand}, nor were they detected by \cite{2017Graham} in 8 other SNe~Ia. The nearby SNe~Ia 2011fe and 2014J were particularly well-studied events \citep{2012Brown,2013Munari,2014Mazzali,2014Foley,2014Goobar,2016Galbany,2016Vallely,2018Dhawan}, but they too showed no hydrogen emission in their late-time spectra \citep{2013Shappee,2015Lundqvist,2016Sand}.

Even SN~2012cg, SN~2012fr, and ASASSN-18bt, events with early excess emission potentially indicative of a single-degenerate progenitor system, did not have hydrogen in their nebular phase spectra \citep{2018Shappee,2017Graham,2018Tucker,2019Dimitriadis_b}. \cite{2016Maguire} looked at a sample of 11 nearby SNe~Ia and found tentative evidence for H$\alpha$ emission in only one event. \cite{2019Sand} examined 8 fast-declining SNe~Ia at nebular phase and could only place upper limits on H$\alpha$ emission. Furthermore, using new and archival spectra of over 100 SNe~Ia, \cite{2019Tucker} found no evidence for the hydrogen or helium emission expected from a non-degenerate companion.

There exists a rare class of thermonuclear SNe that show evidence for interaction with a H-rich circumstellar medium (CSM), the archetype of which is SN~2002ic \citep{2003Hamuy,2004Deng,2004Wang,2004WoodVasey}. Other well-studied ``SNe~Ia-CSM'' include SN~2005gj \citep{2006Aldering,2007Prieto,2008Trundle}, SN~2008J \citep{2012Taddia,2013Fox}, and PTF~11kx \citep{2012Dilday}. \cite{2013Silverman} identified a number of new events and produced the most detailed analysis to date of this class of transients. While H emission signatures are present in these SNe, they are due to interaction with a H-rich CSM and do not constitute detections of stripped companion material \citep{2016Maguire}. These events do not obey the empirical light curve relations (e.g., \citealt{1993Phillips,2006Prieto,2014Burns}) that define ``normal'' SNe~Ia, and they are also considerably brighter (by $\sim1$ mag) than typical SNe~Ia \citep{2013Silverman}.

To date, only one normal Type~Ia SN, ASASSN-18tb \citep{2018BrimacombeATel}, shows compelling evidence for strong H$\alpha$ emission \citep{2019Kollmeier}. Even here, the phenomenon is clearly rare, as it is the only example in the sample of 75 spectra obtained to date for the well-defined 100 Type Ia Supernova sample (100IAS, \citealt{2018Dong100Ias}) and there were none in the larger, heterogeneous sample of \cite{2019Tucker}. ASASSN-18tb was also observed by \textit{TESS} \citep{2015RickerTESS}, providing a high cadence, early-time light curve.  Here we report on these \textit{TESS} observations as well as on additional ground based photometry and spectroscopy.  We describe the observations in \S2, the \textit{TESS} systematics in \S3, the \textit{TESS} early-time light curve in \S4, the spectroscopic characteristics in \S5, and discuss the results in \S6.

\section{Observations}
\label{sec:obs}

\subsection{Discovery and Host Galaxy}
\label{subsec:disc}

ASASSN-18tb (SN 2018fhw) was discovered by the All-Sky Automated Survey for Supernovae \citep[ASAS-SN;][]{2014ShappeeASASSN,2017Kochanek,2017HoloienCat2,2017HoloienCat3,2017HoloienCat1,2019Holoien} in images obtained on UT 2018-08-21.31 (JD 2\,458\,351.81) at J2000 R.A. 04$^\textnormal{h}$18$^\textnormal{m} 06\fs149$ and Decl. $-$63{\degree}36'$56\farcs68$ \citep{2018BrimacombeATel,2018BrimacombeTNS}. From the \cite{2011Schlafly} recalibration of the \cite{1998Schlegel} infrared-based dust map, we find that the supernova suffers relatively little Galactic extinction, $E(B-V)=0.03$ mag.

ASASSN-18tb is located $4\farcs8$ south and $1\farcs9$ east of the
center of 2MASX J04180598--6336523, an extended source in the Two Micron All
Sky Survey \citep[2MASS;][]{2006Skrutskie} with magnitudes $m_J = 15.06 \pm
0.11$, $m_H = 14.23 \pm 0.13$, and $m_K = 13.88 \pm 0.17$. Prior to the
discovery of ASASSN-18tb, there were no public spectroscopic observations of
2MASX J04180598-6336523. However, when obtaining a classification spectrum of
the supernova, \cite{2018Eweis} also obtained a spectrum of the host galaxy.
Using cross-correlations with galaxy templates, they found that it has a
heliocentric redshift of $5090 \pm 30$ km s$^{-1}$ $(z=0.0170 \pm 0.0001)$. This
redshift yields a luminosity distance of 74.2 Mpc assuming $H_0 = 69.6$ km s$^{-1}$ Mpc$^{-1}$,
$\Omega_M = 0.286$, and $\Omega_\Lambda = 0.714$
\citep{2006Wright,2014Bennett}. 

Using the Supernova Identification code
\citep[SNID;][]{TonrySNIDAlgorithm,BlondinSNID}, \cite{2018Eweis} classified
ASASSN-18tb as a spectroscopically normal SN~Ia based on a
SALT spectrum obtained on UT 2018-08-23.3, finding a good match to the Type Ia
SN~2003iv at a phase of $+1$ day beyond maximum light. \cite{2019Kollmeier}
find that, like SN~2003iv \citep{2012Blondin}, ASASSN-18tb has a ``cool''
sub-classification in the scheme of \citet{2006Branch}, and that
photometrically, ASASSN-18tb is a fast-declining, sub-luminous SN~Ia.

\subsection{Photometry}
\label{subsec:phot}

We present photometric observations obtained over the course of 70 days, beginning at MJD~58343 (see Figure~\ref{fig:GroundandSpaceLightCurve}).
Most of the
ground-based observations were obtained using the 1m telescopes and Sinistro
CCDs of the Las Cumbres Observatory Global Telescope Network
\citep[LCOGT;][]{Brown2013}. Additional high-cadence observations near maximum
light were obtained using the quadruple 14-cm ASAS-SN telescopes ``Cassius'' and
``Bohdan Paczy\'{n}ski'' deployed in Cerro Tololo, Chile.
We also present late-time $B$ and $V$ band observations obtained using
the ANDICAM instrument \citep{2003ANDICAM} mounted on the 1.3-m telescope at the Cerro
Tololo Inter-American Observatory (CTIO) operated by the Small \& Moderate Aperture Research Telescope System (SMARTS) Consortium.

ASAS-SN images are processed in an automated pipeline using the \textsc{ISIS}
image subtraction package \citep{1998Alard,2000Alard}. Using the IRAF
\textsc{apphot} package, we performed aperture photometry on the subtracted
images and then calibrated the results using the AAVSO Photometric All-Sky
Survey \citep[APASS;][]{Henden2015}. Reduced images (after bias/dark-frame and flat-field corrections) from LCOGT and the SMARTS 1.3m telescope were downloaded from the respective data archives. We perform point-spread-function (PSF) photometry using the DoPHOT \citep{1993Schechter} package. Optical photometry in the $B$, $V$, $r$, and $i$ bands were calibrated using the APASS standards.

We also obtained images in the $V$, $B$, $U$, $UVW1$, $UVM2$ and $UVW2$ bands
with the \textit{Neil Gehrels Swift Observatory}'s Ultraviolet Optical Telescope
\citep[UVOT;][]{2005RomingUVOT}. The \textit{Swift} UVOT photometry is
extracted using a $5\farcs0$ aperture and a sky annulus with an inner radius
of $15\farcs0$ and an outer radius of $30\farcs0$ with the \textsc{uvotsource}
task in the \textsc{heasoft} package. The \textit{Swift} UVOT photometry is
calibrated in the Vega magnitude system based on the revised zero-points and
sensitivity from \cite{2011Breeveld}.

We characterize the ASASSN-18tb light curve using the \citet{2007Jha} update
of the \citet{1996Riess} and \citet{1998Riess} multicolor light-curve shape
method, \textsc{MLCS2k2}. We find that the peak of the $B$ light curve
occurred at $t_0=58357.33\pm0.12$ MJD.
We reference our observations to this inferred date of maximum light
throughout this work. After accounting for Galactic extinction, we find that extinction from the host galaxy is
negligible. The MLCS2k2 fit yields a light-curve shape parameter $\Delta =
1.41\pm0.03$, squarely in the fast-declining region of parameter space. It has a color stretch of $s_{BV} \approx 0.44$ and $\Delta m_{15} (B)
\approx 2.0$ mag using the relations given by \citet{2018ApJ...869...56B}.
These results are in agreement with those of \citet{2019Kollmeier}, who find
$s_{BV} = 0.50 \pm 0.04$ and $\Delta m_{15}(B) = 2.0 \pm 0.1$ mag using the
SNooPy light curve fitter \citep{2011AJ....141...19B}.
Our MLCS2k2 fits give peak absolute magnitudes for ASASSN-18tb of $M_B = -17.66 \pm 0.09$ and $M_V = -18.05 \pm 0.09$ mag and a slightly closer distance ($65 \pm 4$ Mpc) than the distance of 74.2 Mpc inferred from the redshift. The difference is roughly consistent with the peculiar velocity uncertainty at this redshift \citep[$\sim$0.13 mag, e.g.,][]{2005Reindl}.

Located near the Southern \textit{TESS} continuous viewing zone, ASASSN-18tb was well-observed by \textit{TESS}. This allowed us to extract the Sector 1 and 2 \textit{TESS} light curves that we present in this paper. We used image subtraction \citep{1998Alard,2000Alard} on the full frame images (FFIs) from the first \textit{TESS} data release to produce high fidelity light curves. In principle it is possible to generate a single reference image and then rotate it accordingly for use during multiple sector pointings, but the large pixel scale of the \textit{TESS} observations makes this particularly difficult and introduces a relatively large source of uncertainty.

We instead chose to construct independent reference images for each sector.
The Sector 1 reference image was constructed using 100 FFIs obtained between MJD 58324.8 and 58326.88, and the Sector 2 reference image was constructed using 100 FFIs obtained between MJD 58353.63 and 58355.69.
In each case these are the first 100 FFIs obtained during the sector. The light curves change little when different images are used to build the reference, and our light curves are consistent with those obtained using the public \textit{TESS} aperture photometry tool \textsc{eleanor}\footnote{\url{https://adina.feinste.in/eleanor/}} \citep{2019Feinstein}.

Because the Sector 2 reference was constructed from images containing a considerable amount of flux from the supernova, fluxes in the raw difference light curve for Sector 2 are systematically lower than the intrinsic values. We correct for this by using a power-law fit (described in more detail in Section~\ref{sec:earlyLC}) to align the Sector 1 and 2 light curves. The offset is calculated by fitting the first day (48 epochs) of Sector 2 photometry to the best-fit single-component power-law for the Sector 1 photometry.

The Sector 1 and 2 fluxes were converted into \textit{TESS}-band magnitudes by adopting a zero point of 20.44 electrons per second in the FFIs, based on the values quoted in the TESS Instrument Handbook\footnote{\url{https://archive.stsci.edu/missions/tess/doc/TESS_Instrument_Handbook_v0.1.pdf}}. \textit{TESS} observes in a single broad-band filter, ranging from about 6000--10000\,\AA{} with an effective wavelength of $\sim$8000 \AA{}, and the \textit{TESS} magnitude system is calibrated to the Vega system \citep{2015SullivanTESS}.

The complete photometry is shown in  Figure~\ref{fig:GroundandSpaceLightCurve}, and all of the data is available in machine-readable format in the online version of the paper.

\begin{figure}
\centering
\includegraphics[width=\columnwidth]{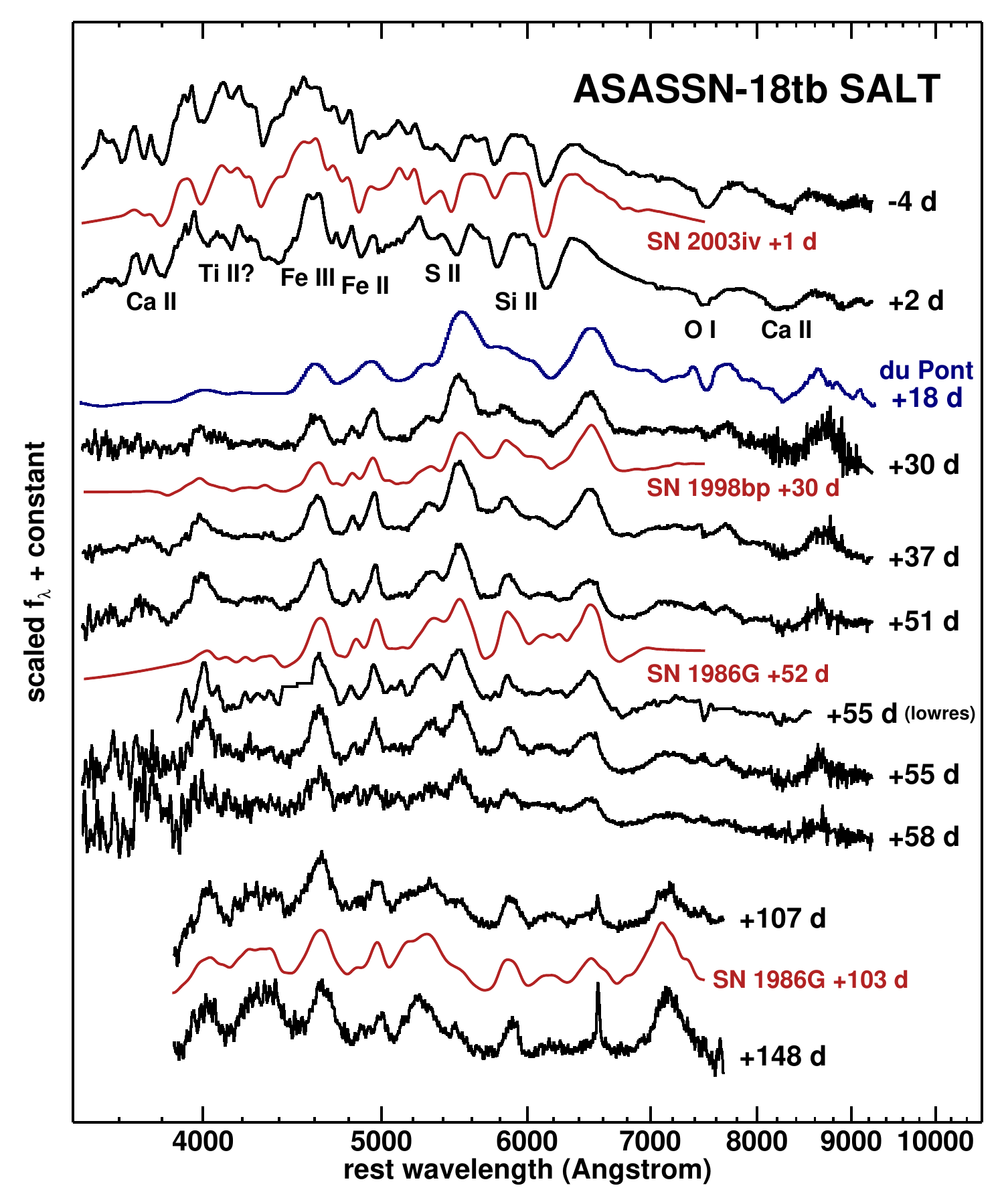}
\caption{SALT and du Pont spectra of ASASSN-18tb obtained at phases ranging from
pre-maximum light to the early nebular phase. Note the clear presence of
H$\alpha$ emission in the nebular phase spectra. Also shown for comparison
are spectra of the
fast-declining SNe~Ia 2003iv, 1998bp, and 1986G
\citep{2001Richardson,2002Hamuy,2012Blondin}. The ASASSN-18tb spectra have been
smoothed using a Savitzky-Golay filter for presentation.}
\label{fig:SALTSpectra}
\end{figure}

\begin{figure*}
\includegraphics[width=0.49\textwidth]{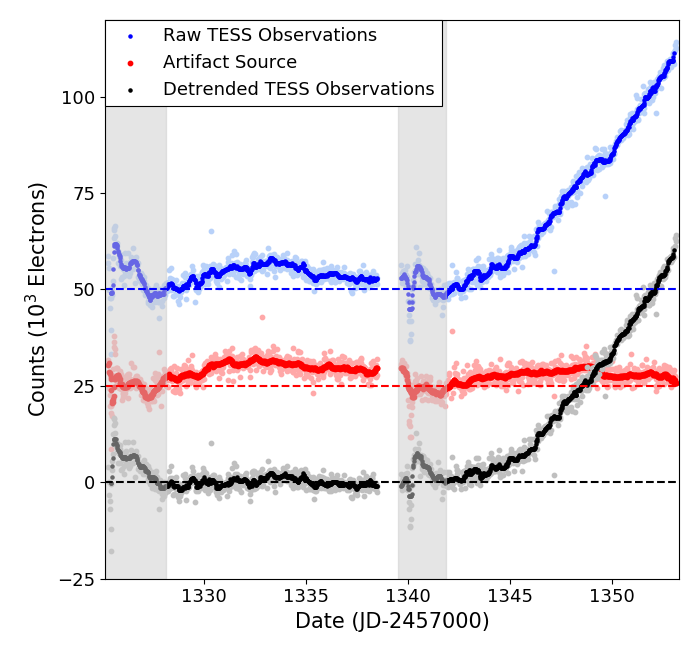}
\hfill
\includegraphics[width=0.49\textwidth]{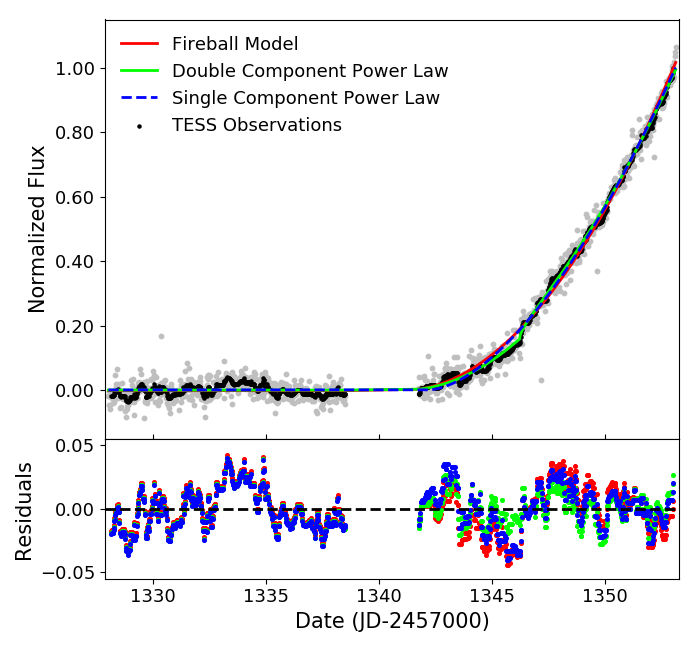}
\caption{The \textit{TESS} Sector 1 light curve of ASASSN-18tb obtained using image subtraction. The raw and detrended light curves of ASASSN-18tb are shown on the left, as well as the artifact-tracing light curve of Stars 1 and 2 (see text for details). Flux values for every epoch are shown in lighter colors, and a 6 hour rolling median of these flux values is shown in darker colors. Vertical gray regions indicate times when considerable scattered Earthshine artifacts are present. The detrended \textit{TESS} light curve of ASASSN-18tb as well as three simple power-law fits and their residuals are shown on the right. Normalized flux is given relative to the maximum Sector 1 flux of 0.701 mJy. Although the rise is shallower (with index $1.69\pm0.04$) than that of a simple expanding fireball model, there is no compelling evidence for additional structure beyond a single-component power law.}
\label{fig:EarlyLC}
\end{figure*}

\begin{table*}
\begin{centering}
\caption{Power-law Fits.}
\label{tab:fitparams}
\begin{tabular}{ccccccccc}
\hline\hline
Model & $z$ ($\mu$Jy) & $t_1$ (MJD) & $h_1$ ($\mu$Jy) & $a_1$ & $t_2$ (MJD) & $h_2$ ($\mu$Jy) & $a_2$ & $\chi^2/\nu$ \\
\hline\hline
Fireball & $-0.6\pm0.6$ & $58340.48 \pm 0.06$ & $4.82\pm0.06$ & $\equiv2$ & $\cdots$ & $\cdots$ & $\cdots$ &  0.995 \\
Single & $+0.9\pm0.9$ & $58341.68 \pm 0.16$ & $12.3 \pm 1.50$ & $1.69 \pm 0.04$ & $\cdots$ & $\cdots$ & $\cdots$ &  1.007 \\
Double & $+0.6\pm0.6$ & $58340.61 \pm 0.37$ & $4.32 \pm 1.41$ & $1.99 \pm 0.10$ & $58345.73 \pm 0.12$ & $36.23 \pm 5.56$ & $0.45 \pm 0.11$ & 1.003 \\
\hline\hline
\end{tabular} \\
\end{centering}
\end{table*}

\subsection{Spectroscopy}
\label{subsec:spec}

The bulk of the spectra we present in this paper were obtained using the Southern
African Large Telescope (SALT) with the Robert Stobie Spectrograph
\citep{2006Buckley}. We used the PG0900 grating with a $1\farcs5$ slit
at multiple tilt positions to cover the optical wavelength range with a typical
resolution of $R\sim1000$. The total exposure time varied from 1932 to 3000
seconds as the supernova faded. Our first spectrum provided the classification reported by \cite{2018Eweis} and was obtained on UT
2018-08-23, four days before ASASSN-18tb attained maximum light. Our last spectrum was taken
on UT 2019-01-25, nearly 150 days after maximum light. The SALT spectra
were reduced using a custom pipeline based on the PySALT package
\citep{2010Crawford}, which accounts for basic CCD characteristics (e.g.,
cross-talk, bias and gain correction), removal of cosmic rays, wavelength
calibration, and relative flux calibration. Standard IRAF/Pyraf routines were
used to accurately account for sky and galaxy background removal.

The 1D spectra are delivered with a nominal dispersion of $\sim 1$\AA{}/pixel. For our analysis in \S\ref{sec:spectra}, each spectrum is rebinned to 7\AA{}/pixel which is the approximate spectral resolution at \Ha. The RMS of the original pixels within each bin is used to estimate the uncertainty at each wavelength in the rebinned spectrum. To model the continuum we use a 2nd-order Savitsky-Golay polynomial of variable width. The continuum fit width is varied from $2\,000-5\,000~\rm{km}~\rm s^{-1}$ at each pixel, and we take the median of these values as the continuum level and the RMS as the uncertainty.

Because of the instrument design, which has a moving, field-dependent and
under-filled entrance pupil, observations of spectrophotometric flux standards
do not suffice to provide accurate absolute flux calibration for SALT
observations (see, e.g., \citealt{2018Buckley}). Therefore, in order to
characterize the interesting H$\alpha$ signature in our spectra, we
recalibrate our observed spectra to match the measured photometry using a
low-order polynomial in wavelength. Fortunately, we have contemporaneous LCOGT $BVri$ coverage for most of our spectroscopic
epochs and can perform the
absolute flux calibration reasonably well ($\pm$ 5\% estimated uncertainty).
For the three late-phase spectra obtained beyond MJD 58420, we do not have sufficient multi-filter coverage from our
photometric observations, so we estimate $BVri$ from extrapolations of the
\textsc{MLCS2k2} fits. In this regime, we assume our flux calibration error is
of order $\pm 10\%$.


We also obtained one lower-resolution spectrum with SALT on UT 2018-10-22 using the PG0300 grating and the same 1$\farcs$5 slit, yielding $R \approx 350$ in a 1600 second exposure at one grating tilt position. We further observed ASASSN-18tb with the B\&C spectrograph on the Ir\'{e}n\'{e}e du Pont telescope at Las Campanas Observatory on UT 2018-09-14 using the 300 line grating with the 150 \micron~slit in three 1000 second exposures. These spectra were reduced using standard routines, and recalibrated to match the photometry as described above. Because of the lower spectral resolution, we have not used these spectra in the further analysis described below.  Figure \ref{fig:SALTSpectra} shows the spectral evolution of ASASSN-18tb with other fast-declining SNe~Ia for comparison.

\section{TESS Systematics}
\label{sec:systematics}

\begin{figure}
\includegraphics[width=\columnwidth]{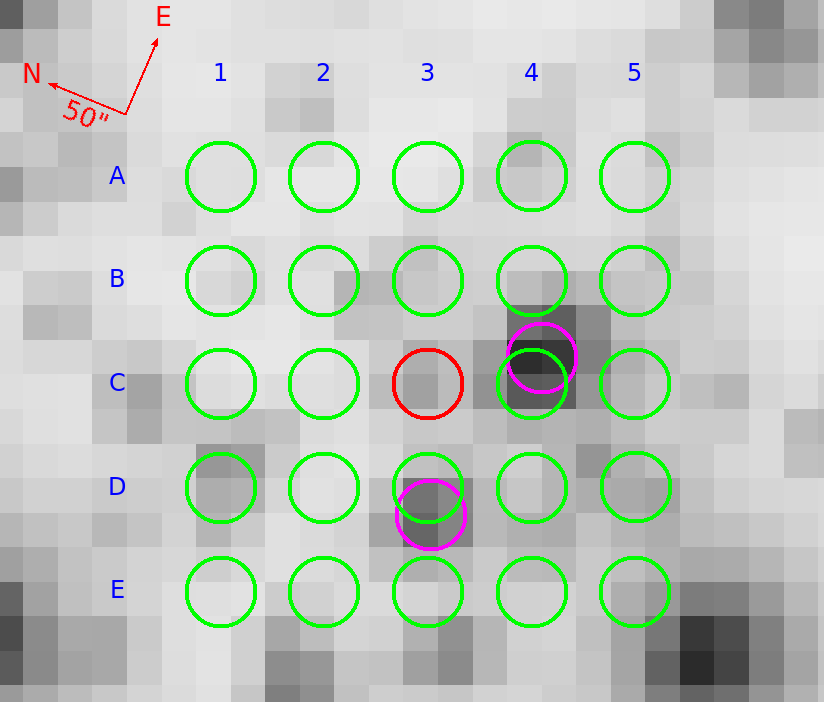}
\caption{The $5\times5$ grid of points for which we obtain image subtraction light curves. The image is the median combination of the last 100 \textit{TESS} Sector 1 FFIs, when the SN is brightest in Sector 1. The grid points are spaced 3 \textit{TESS} pixels away from one another ($\sim 63''$). The red circle at the center of the grid (point C3) indicates the location of ASASSN-18tb, and the two magenta circles located near points C4 and D3 indicate the locations of Star 1 and Star 2, respectively, the stars we consider as tracers of artifacts in the raw supernova light curve.}
\label{fig:CheckFinder}
\end{figure}

\begin{figure*}
\includegraphics[width=0.90\textwidth]{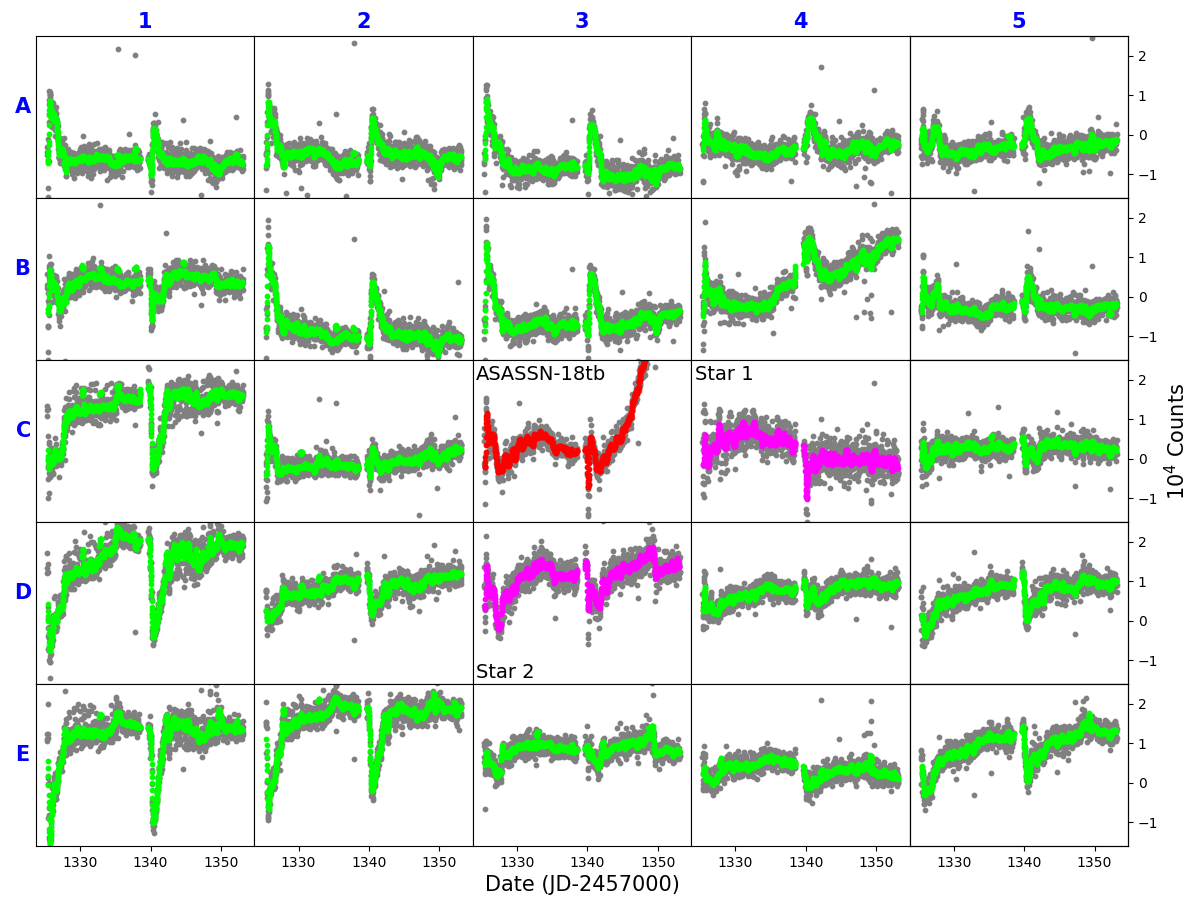}
\caption{Image subtraction light curves obtained for a $5\times5$ grid of points centered on the location of ASASSN-18tb. These light curves correspond to the test coordinates indicated in Figure~\ref{fig:CheckFinder}. Flux values for every epoch are shown in gray, and a 6 hour rolling median of these flux values is shown in color. We have replaced the C4 and D3 light curves with those obtained for the exact locations of Star 1 and Star 2, respectively. Note how well Star 2 traces the pre-explosion bump artifact in the raw ASASSN-18tb light curve.}
\label{fig:CheckGrid}
\end{figure*}

The raw \textit{TESS} Sector 1 image subtraction light curve of ASASSN-18tb is shown in blue in the left panel of Figure~\ref{fig:EarlyLC}. It is clear by inspection that there are a number of systematic artifacts present in the data, some of which are fairly well understood and discussed in the TESS Instrument Handbook and TESS Data Release Notes\footnote{\url{https://archive.stsci.edu/tess/tess_drn.html}}. For instance, we observe high frequency $\sim24$ hour oscillations in the light curve that are likely introduced in the image backgrounds by the rotation of the Earth, as discussed in Section~1.3 of the Sector 1 TESS Data Release Notes\footnote{\url{https://archive.stsci.edu/missions/tess/doc/tess_drn/tess_sector_01_drn01_v01.pdf}}.

While these low-level oscillations are not significant for the scientific goals of our analysis, some of the other systematics present in the data are. Section~7.3.2 of the TESS Instrument Handbook discusses the presence of a patch of scattered Earthlight in TESS FFIs whose structure and intensity depends on the Earth elevation, azimuth, and distance. We visually inspected the TESS Sector 1 FFIs and found that the brightest component of this patch is spatially coincident with ASASSN-18tb for $\sim3$ days at the start of each orbit. No straightforward method exists to account for this artifact in our reductions, so we exclude these artifact regions from our analysis. These exclusion windows are shown as the gray shaded regions in the left panel of Figure~\ref{fig:EarlyLC}.

The raw light curve suggests the presence of significant pre-explosion emission for $\sim10$ days prior to the supernova. Such a signature is entirely unprecedented for SNe~Ia, both theoretically and observationally, so we put some effort into investigating its origin. To do so, we obtained image subtraction light curves for a $5\times5$ grid of test coordinates surrounding ASASSN-18tb using the same reference image (shown in Figure~\ref{fig:CheckFinder} overlayed on a sample FFI). The sources were spaced 3 \textit{TESS} pixels away from one another to ensure sampling on a large spatial scale. The resulting test light curves are shown in Figure~\ref{fig:CheckGrid}. We note that the light curves obtained for the test coordinates located at C4 and D3 show a similar bump artifact.

C4 and D3 lie very near the positions of two relatively bright stars in the Gaia Data Release 2 catalog \citep{2016Gaia,2018Gaia,2018ArenouGaia}: Gaia DR2 4676041915767041280 $(G_{RP}=12.3438\pm0.0005 \textnormal{ mag})$ and Gaia DR2 4676043427595528448 $(G_{RP}=13.653\pm0.001 \textnormal{ mag})$, respectively. Here we will simply refer to the two stars as Star 1 and Star 2, where Star 1 is the star nearest to point C4 (Gaia DR2 4676041915767041280) and Star 2 is the star nearest to the point D3 (Gaia DR2 4676043427595528448). The mean combination of the Star 1 and Star 2 light curves traces the early-time artifact structure in the ASASSN-18tb light curve better than either of the individual light curves. When either of the star light curves is used by itself, the pre-explosion light curve shows a small linear residual. As can be seen in Figure~\ref{fig:EarlyLC}, this is not the case when the mean combination is used.


We thus use the mean combination of the Star 1 and Star 2 light curves in order to correct for systematics. This is shown as the red artifact light curve in Figure~\ref{fig:EarlyLC}. We subtract this artifact light curve from the raw ASASSN-18tb light curve, emphasizing that we apply no multiplicative factor to match the scale of the bump artifact in the two light curves. That the two light curves exhibit nearly identical structure prior to explosion further confirms that the bump is an artifact and not intrinsic to the supernova. After removing the bump artifact, we force the flux zeropoint of the detrended light curve to the average value of the observations obtained from MJD 58327.5 to 58338, corresponding to an average of all observations obtained over the first Sector 1 orbit after removing the window where the bright patch of scattered Earthlight is spatially coincident with ASASSN-18tb.

\section{Early Light Curve}
\label{sec:earlyLC}

\begin{figure}
\centering
\includegraphics[width=\columnwidth]{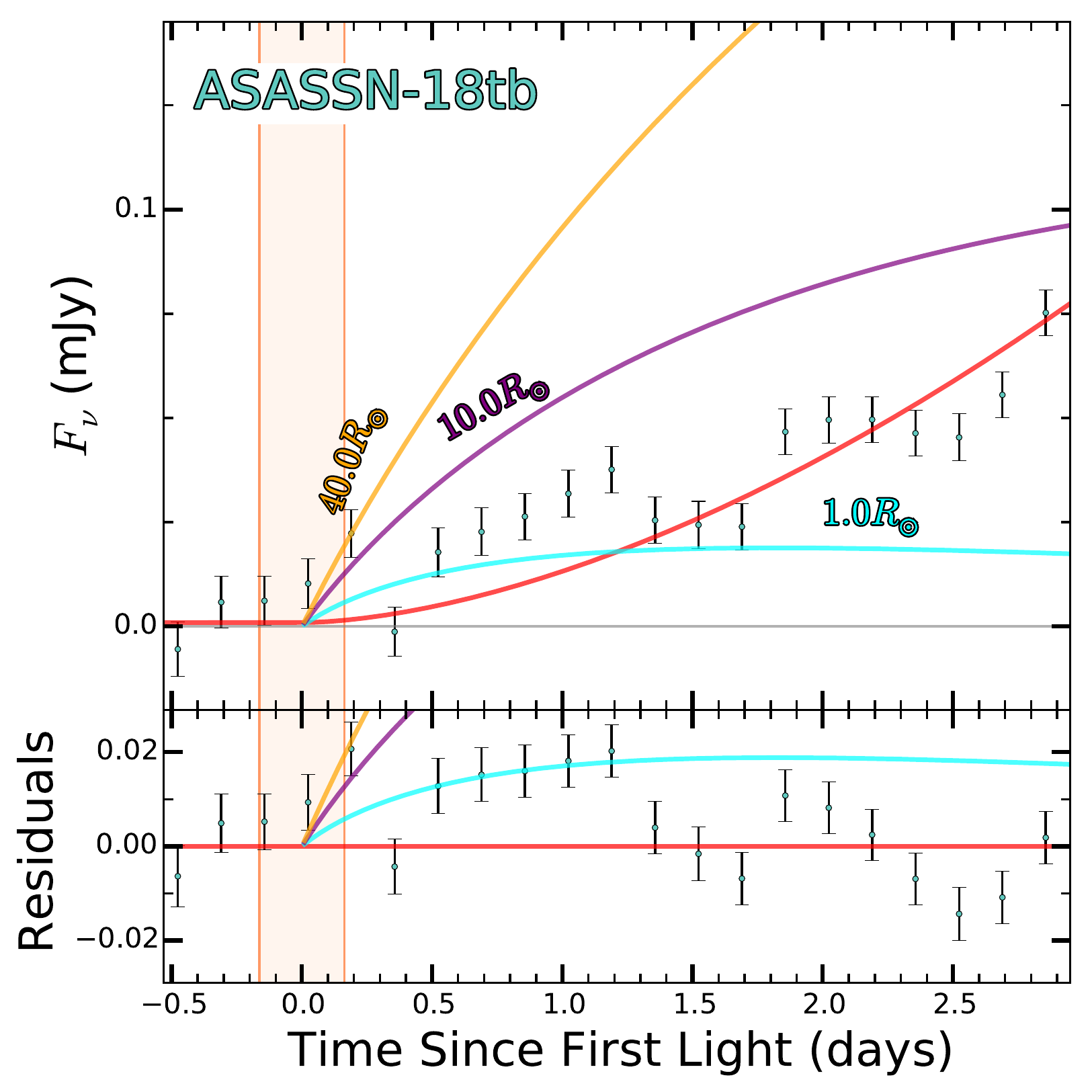}
\caption{The early light curve of ASASSN-18tb compared to the companion interaction models from \protect\cite{2010Kasen}. We adopt $t_1$ from our best-fit single-compnent power-law model as the time of first light, and \textit{TESS} data is shown for 4 hour bins. Our best-fit single-component power-law model is shown in red, and interaction models for non-degenerate $1~R_\odot$, $10~R_\odot$, and $40~R_\odot$ companions are shown in cyan, purple, and gold, respectively. These models are for a viewing angle $(\theta=45\degree)$ where the predicted effect is strong. The lower panel shows residuals relative to our best-fit single-component power-law model.}
\label{fig:InteractionModels}
\end{figure}

The \textit{TESS} light curve of ASASSN-18tb, after accounting for the systematics described in Section~\ref{sec:systematics}, is shown in the right panel of Figure~\ref{fig:EarlyLC}. It is also provided in machine-readable format in the online version of the paper. The light curve does not show evidence of the double-component rise observed for ASASSN-18bt by $K2$ \citep{2019Shappee,2019Dimitriadis}, but motivated by the identification of strong H$\alpha$ emission in the spectra of this supernova, we fit a number of power-law models to the light curve in order to better characterize its properties. The light curve uncertainties were estimated by measuring the root-mean-square scatter $\sigma$ of the pre-explosion observations obtained between MJD 58327.5 and 58338.

The simplest of these models is the expanding fireball model, $f=z+{h_1}(t-t_1)^2$, with three parameters. The fireball model assumes a homologously expanding ejecta, which determines the temporal exponent \citep{Riess1999,Nugent2011}. We also fit arbitrary index power-law models of the form
\begin{flalign}
&f= z \textnormal{ when } t<t_1,&\\
&f= z + h_1(t-t_1)^{\alpha_1} \textnormal{ when } t_1<t<t_2,&\\
&f= z + h_1(t-t_1)^{\alpha_1} + h_2(t-t_2)^{\alpha_2} \textnormal{ when } t_2<t,&
\end{flalign}
where we obtain a single-component fit by simply fixing $h_2\equiv0$. The single-component power-law model thus has four parameters, while the double-component power-law model has seven. We fit these models using the \textsc{scipy.optimize.curve\_fit} package's Trust Region Reflective method, and our best-fit models are shown in the right panel of Figure~\ref{fig:EarlyLC}.

Our best-fit fireball model is shown using the solid red curve. The fit is quite reasonable, although the model has moderate discrepancies with the observed flux. Our best-fit single- and double-component models are shown using the dashed blue and solid green lines, respectively. By eye, the three models are virtually indistinguishable, indicating that there is no need to invoke a model more complex than the single-component power-law. The $\chi^2$ per degree of freedom $(\nu)$ for each of our three fits are given in Table~\ref{tab:fitparams}. We find no evidence to justify using the double-component power law, as it produces no significant change in $\chi^2/\nu$.

Although these simple fits indicate that there is no significant secondary source of early-time emission, it is worth examining the potential contribution to the light curve from ejecta interaction with a non-degenerate companion given the prominent H$\alpha$ emission seen in the spectra. \cite{2010Kasen} showed that such interaction would produce significant additional flux for certain viewing angles and provided concise analytic solutions, which we utilize here. Interaction models for $1~R_\odot$, $10~R_\odot$, and $40~R_\odot$ companions are shown in Figure~\ref{fig:InteractionModels}, along with our best-fit single-component power-law and the \textit{TESS} observations immediately surrounding the beginning of the explosion.

These interaction models depend strongly on the viewing angle. One generally only expects to see significant signal for viewing angles looking down on the collision region $(\theta \sim 0\degree)$, and the models in Figure~\ref{fig:InteractionModels} assume a value fairly close to this optimal viewing angle $(\theta=45\degree)$. Under this assumption, it is clear that we can rule out any companion significantly larger than $1~R_\odot$ for ASASSN-18tb. However, constraining the viewing angle for any individual event is extremely difficult. In practice, one could almost completely mask the interaction signature from even a massive star if it were viewed at an angle of $\theta\sim180\degree$. 

We note, however, that the H$\alpha$ emission we observe is slightly blue-shifted (See Section~\ref{sec:spectra}). If the H$\alpha$ signature is indeed produced by swept-up material from a companion star, it would appear blueshifted only for viewing angles relatively close to $\theta\sim0\degree$ \citep{2018Boty}. We thus regard the optimal viewing angle models shown in Figure~\ref{fig:InteractionModels} as instructive, if not definitive, and consider non-degenerate companions of $R \gtrsim R_\odot$ to be inconsistent with our observations.

\cite{2019Kollmeier} showed that ASASSN-18tb is a broadly normal under-luminous SN~Ia based on its empirical characteristics. Here we use our excellent set of photometric observations to estimate the near-maximum bolometric luminosity and examine the physical parameters needed to produce it. As in \cite{2018Vallely}, we estimate the bolometric luminosity using Markov Chain Monte Carlo (MCMC) methods to fit a blackbody to the observed spectral energy distribution. We limit our analysis to near-maximum dates with good filter coverage $(n_{filters}>4)$. All photometry was corrected for Galactic foreground extinction prior to being fit. As discussed in Section~\ref{subsec:phot}, there is no evidence for additional extinction from the host galaxy. 

Semi-analytic models for light curves powered by the radioactive decay of $^{56}$Ni have been available for some time \citep{1979Arnett,1982Arnett}. We can estimate the ejecta mass $(M_{ej})$ by assuming that the light curve peaks approximately at the diffusion time $(t_d)$. The ejecta mass is then approximated by
\begin{equation}
M_{ej}=t_d^2\frac{4\pi c v_{ej}}{3\kappa} \approx \Big( \frac{t_{peak}-t_1}{1+z} \Big) ^2\frac{4\pi c v_{ej}}{3\kappa},
\end{equation}
where $c$ is the speed of light, $M_{ej}$ is the ejecta mass, $\kappa$ is the opacity of the ejecta, $v_{ej}$ is the ejecta velocity, $z$ is the redshift, $t_{peak}$ is the time of maximum light, and $t_1$ is the time of explosion. We will assume an approximate 10\% systematic uncertainty for our ejecta mass estimate (see, e.g., \citealt{2013Blondin}, \citealt{2018Wilk}, and \citealt{2006Stritzinger}).

We again adopt MJD 58357.33 for $t_{peak}$, and we use the $t_1$ value from our
single-component power-law model (MJD 58341.68). For the ejecta velocity, we adopt $v_{ej}=10\,000$ km s$^{-1}$, consistent with estimates of the expansion velocity from \cite{2018Eweis}. Like \cite{2018Khatami} and \cite{2018Sukhbold}, we adopt an opacity value $\kappa=0.1$ cm$^2$ g$^{-1}$ for our model that is typical of ionized ejecta \citep{2000Pinto,2017Arnett,2017BranchWheeler}. Using these values, we obtain a slightly sub-Chandrasekhar ejecta mass of $M_{ej}=1.11\pm0.12~\textnormal{M}_\odot$. This is consistent with the results of \cite{2019Scalzo}, who found a preference for sub-Chandrasekhar mass explosions among 1991bg-like SNe~Ia.

We can estimate the amount of $^{56}$Ni $(M_{Ni})$ synthesized in the explosion using Arnett's rule, noting that at time $t_d$ after explosion, when the supernova attains maximum brightness, the luminosity will approximately equal that of the instantaneous radioactive decay power from the $^{56}$Ni$\rightarrow^{56}$Co$\rightarrow^{56}$Fe decay chain. We can then solve for $M_{Ni}$ as
\begin{equation}
M_{Ni} = \frac{L_{peak}}{\textnormal{ergs s}^{-1}} \big( C_{Ni}e^{-t_d/\tau_{Ni}} + C_{Co}e^{-t_d/\tau_{Co}} \big)^{-1} \textnormal{ M}_\odot,
\end{equation}
where the decay times of $^{56}$Ni and $^{56}$Co are known to be $\tau_{Ni}$=8.77 days and $\tau_{Co}$=111.3 days, respectively \citep{2005Stritzinger,2006Taubenberger,1987Martin}, and $C_{Ni} \approx 6.45\times10^{43}$ and $C_{Co} \approx 1.45\times10^{43}$ \citep{1994Nadyozhin,2018Sukhbold}.

After including a 4 Mpc uncertainty in our redshift-estimated luminosity distance ($74.2\pm4$ Mpc), our MCMC fit yields $L_{peak}=(7.4\pm1.1)\times10^{42}$ ergs s$^{-1}$.
From this we find that nominally $M_{Ni}=0.31\pm0.05~\textnormal{M}_\odot$.
However, this simple model is typically only accurate to within 20\% \citep{2013Blondin,2017Hoeflich}, and it tends to overestimate $M_{Ni}$ for sub-luminous SNe~Ia like ASASSN-18tb \citep{2018Khatami}.
To account for this, we report a lower limit that is 20\% smaller than that of the nominal $M_{Ni}$ estimate. We thus find that $M_{Ni}=0.21\textnormal{ -- }0.36~\textnormal{M}_\odot$. This is comparable to the $M_{Ni}$ estimates found by \cite{2019Scalzo} for the 1991bg-like SNe 2006gt and 2007ba, but is somewhat larger than the $M_{Ni}\sim0.1\textnormal{M}_\odot$ estimated for SN~1991bg itself by \cite{2006Stritzinger}. Futhermore, it is reasonably consistent with the $M_{Ni}\approx0.2~\textnormal{M}_\odot$ estimate obtained using the \cite{2018Goldstein} fitting functions calibrated using a library of radiative transfer models.

\section{Early- and Late-phase Spectroscopy}
\label{sec:spectra}

\begin{figure*}
    \centering
    \includegraphics[width=\linewidth]{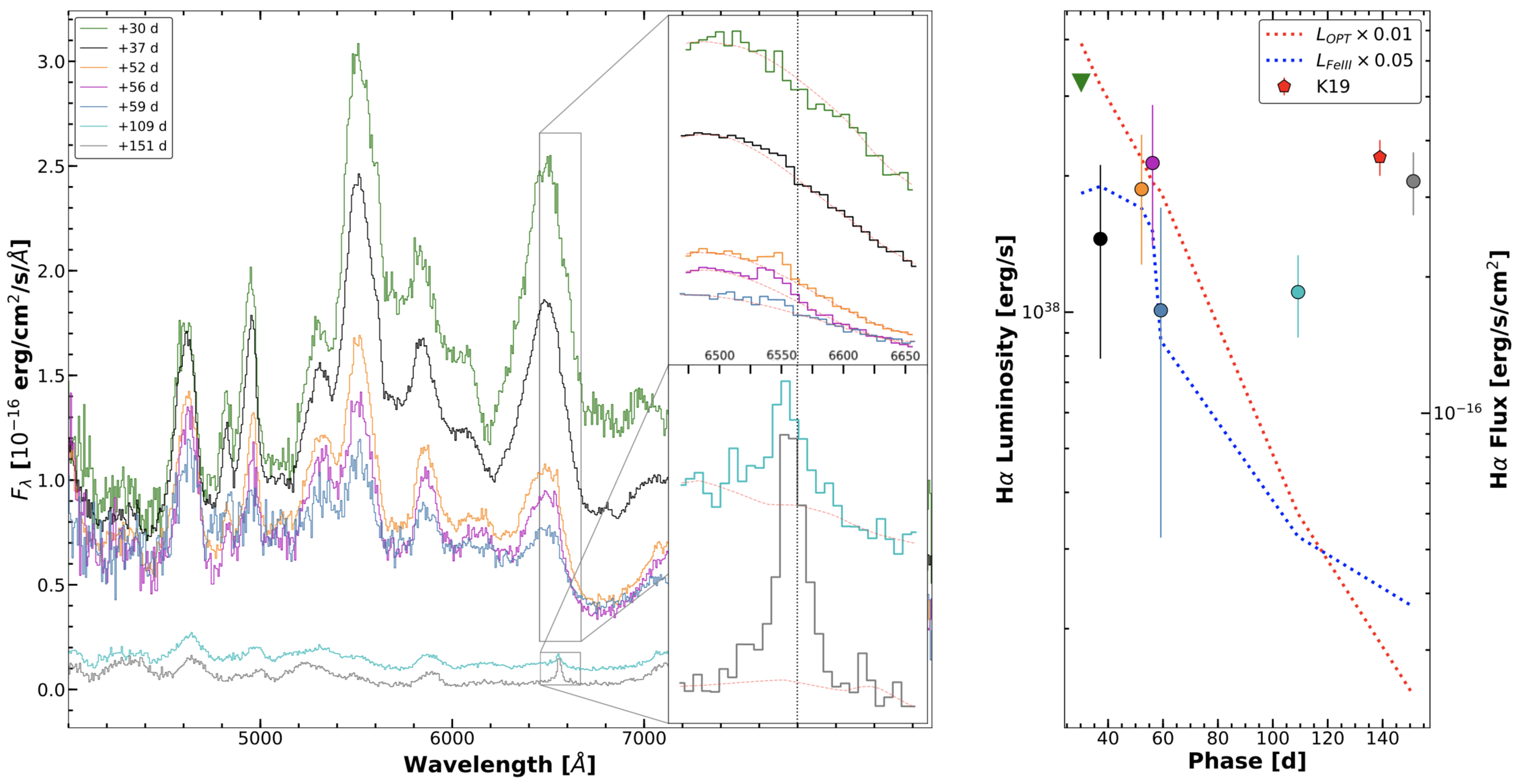}
    \caption{
    \textit{Left}: Flux-calibrated spectral evolution of \name. \textit{Insets}: Zoom-in view around \Ha for early-phase (top) and late-phase (bottom) spectra. The dashed red lines indicate the continuum fit and the vertical dotted line indicates the rest wavelength of \Ha. \textit{Right}: Evolution of the integrated \Ha luminosity as a function of time from peak brightness compared to the evolution of the approximate optical luminosity (integrated from $4000-7000$~\AA{}, red) and the integrated Fe~III$~\lambda4660$ luminosity (blue) over the same timespan. Colored points correspond to the same color spectrum in the left panel. The right axis denotes the measured flux values since the distance is uncertain. K19 refers to values taken from \citet{2019Kollmeier}. The \Ha luminosity does not track the falling luminosity of the SN and is consistent with a constant value.
    }
    \label{fig:spec-evol}
\end{figure*}

\begin{figure}
    \centering
    \includegraphics[width=\linewidth]{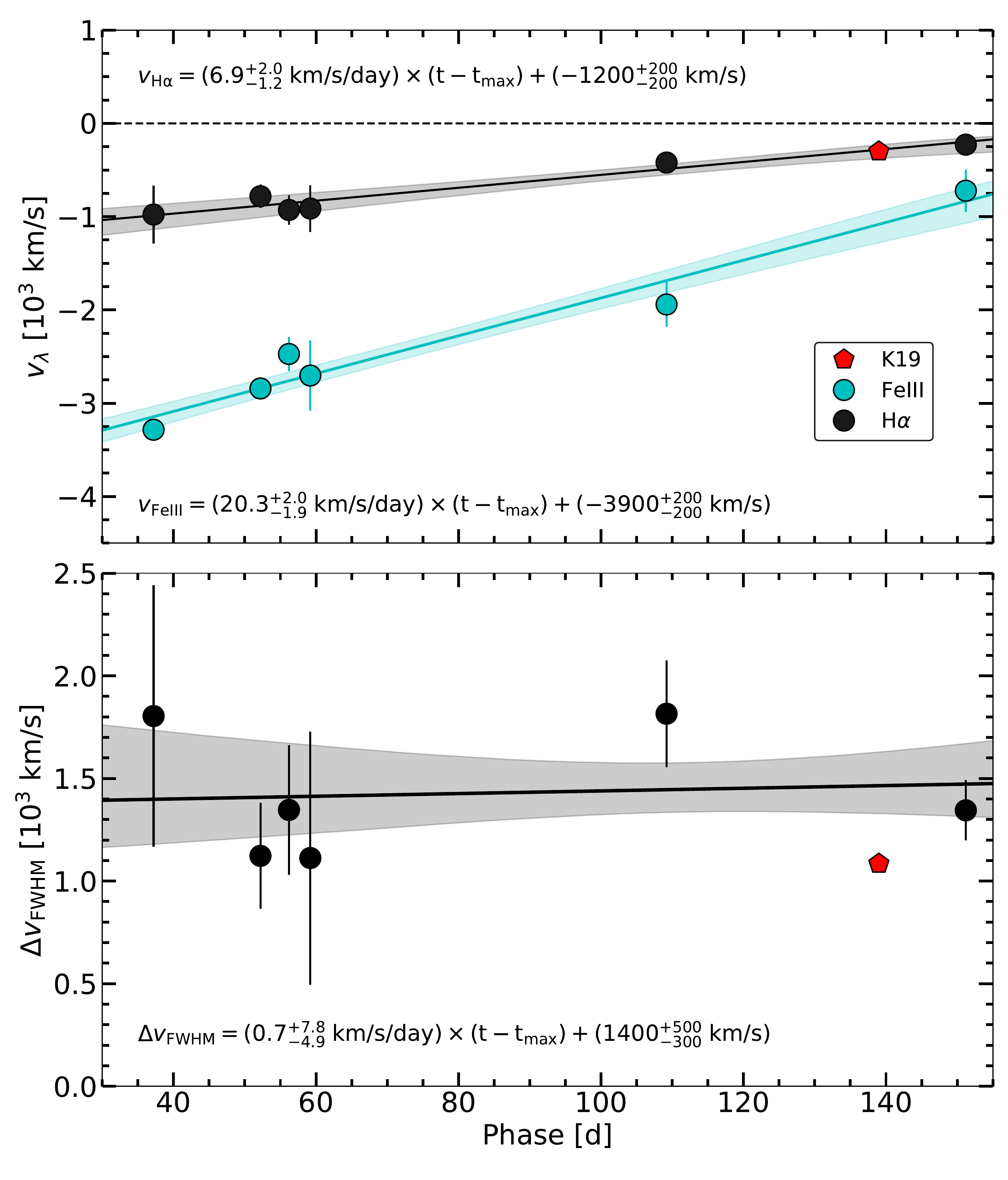}
    \caption{Evolution of the \Ha profile as a function of time. K19 refers to values taken from \citet{2019Kollmeier} which are not included in the fitting process due to the lack of reported uncertainties.
    \textit{Top}: Velocity shift as a function of time for the \Ha (black) and the Fe~III$~\lambda4660$ (blue) emission lines. Note that our best-fit solution for \Ha aligns with the K19 measurement even though it was omitted from the fit. 
    \textit{Bottom}: Width of the \Ha emission feature as a function of time. The width is roughly constant over the span of our spectral observations.
    }
    \label{fig:linefits}
\end{figure}

Our SALT spectra span $-4$ to $+148$ days relative to maximum light. As shown in
Figure \ref{fig:SALTSpectra}, excluding the broad \Ha emission, the spectra of
\name share many qualities with the under-luminous 91bg-like class of
thermonuclear supernovae \citep{1992Fillippenko,1993Leibundgut,1994Hamuy}.
The near-maximum spectra exhibit the Si~II absorption feature typical of
\sne plus hints of the Ti~II absorption of 91bg-like objects. Additionally,
broad [Ca~II] emission at $\lambda\sim 7300$\AA{} is present in the late-phase spectra.

The most intriguing aspect of \name  is the presence of broad, FWHM $\sim 1000~\rm{km}~\rm s^{-1}$ \Ha emission \citep{2019Kollmeier}. While the \Ha emission is clearly visible in the $> +100~\rm d$ late-phase spectra (Fig. \ref{fig:SALTSpectra}), we also see evidence of \Ha emission in spectra starting roughly $+50~\rm d$ after peak light (Fig. \ref{fig:spec-evol}). There is a tentative detection in the $+37~\rm d$ spectrum, and a non-detection in the $+30~\rm d$ spectrum. The upper limit on \Ha for the +30~d spectrum assumes an \Ha profile similar to the one detected in the +37~day spectrum, using a FWHM velocity of $\sim 1500~\rm{km}~\rm s^{-1}$ blue-shifted by $\sim 1000~\rm{km}~\rm s^{-1}$.

To characterize the nature of the \Ha emission, we subtract off the continuum and fit the emission line with a Gaussian profile. Fig. \ref{fig:linefits} shows the line center and FWHM evolution of the \Ha emission. For comparison, we also include the evolution of the Fe~III$~\lambda4660$ line in the top panel. The line center of the Fe~III feature is measured by fitting a Gaussian profile plus linear continuum at each epoch. 

We use a linear model to calculate the temporal evolution for each line,
\begin{equation}
    v_\lambda = \dot{v}_\lambda (t-t_{\rm{max}}) + b_\lambda.
\end{equation}

\noindent Here, $v_\lambda$ is the velocity shift from rest for the \Ha ($v_{\rm H\alpha}$) and Fe~III ($v_{\rm{FeIII}}$) lines at phase ($t-t_{\rm{max}}$) days. The values $\dot{v}_\lambda$ and $b_\lambda$ are computed using linear least-squares fitting and a bootstrap-resampling technique to estimate the uncertainties. The value from \citet{2019Kollmeier} does not have a reported uncertainty, so we do not include it in our fit. We see clear evidence for varying line velocities, with $\dot{v}_{\rm H\alpha} = 6.9^{+2.0}_{-1.2} ~\rm{km~s^{-1}/day}$ and $\dot{v}_{\rm{FeIII}} = 20.3^{+2.0}_{-1.9} ~\rm{km~s^{-1}/day}$. The line velocity drift rates for these two lines are discrepant at $\sim 5 \sigma$.

We also fit the FWHM velocity ($\Delta v_{\rm{FWHM}}$) of the \Ha emission, shown in the bottom panel of Fig. \ref{fig:linefits}, and find a weighted mean of $\Delta v_{\rm{FWHM}} = 1390\pm220~\rm{km}~\rm s^{-1}$. To determine the temporal evolution of $\Delta v_{\rm{FWHM}}$, we use the same linear model and bootstrap-resampling technique as before, finding the width of the \Ha emission is consistent with a temporally constant value, although the uncertainties are large (Fig. \ref{fig:linefits}).

Because a roughly constant \Ha flux is consistent with CSM SNe Ia, we searched for other emission lines associated with circumstellar interaction, such as He~I and H$\beta$. No other broad emission lines atypical of \sne are found in our spectra and we place upper limits on these non-detections. For the early-phase spectra ($<100~\rm d$ after max), we place a limit on the Balmer decrement of $F_{\Ha}/F_{\rm H\beta}\gtrsim 2$. For the late-phase spectra we place a limit of $F_{\Ha}/F_{\rm H\beta}\gtrsim 5$, consistent with the value found by \citet{2019Kollmeier}. For the non-detection of He~I$~\lambda5875$, we find $F_{\Ha}/F_{\rm{He\texttt{I}}}\gtrsim 3$ for all spectra.

\section{Discussion and Conclusions}
\label{sec:conclusion}
ASASSN-18tb is clearly an unusual event, and its place in the ever-changing menagerie of supernova taxonomy will likely be the source of ongoing discussion. The detection of strong H$\alpha$ emission in an empirically normal SN~Ia is unprecedented. \cite{2019Kollmeier} discussed possible sources for this signature, including swept up material from a non-degenerate companion and CSM interaction, but their analysis was necessarily limited by having only one post-maximum spectrum to examine. With our additional photometric and spectroscopic observations, we can provide a more extensive discussion of the origin of ASASSN-18tb and its unusual characteristics.

While the $\textnormal{FWHM}\sim1000$ km s$^{-1}$ H$\alpha$ emission we observe in the late-time spectra is consistent with the predicted signatures of swept up material from a non-degenerate companion, other aspects of the emission are not. It is difficult to reconcile the approximately constant H$\alpha$ luminosity with this interpretation, as one would expect the H$\alpha$ to follow the SN bolometric luminosity. This is because the H$\alpha$ emission is powered by gamma-ray deposition from the decay of $^{56}$Ni, the same source which powers the SN light curve \citep{2018Boty}. Additionally, if the H were swept up in the SN ejecta the velocity evolution of H$\alpha$ emission should approximately trace that of Fe~III \citep{2018Boty}, but we do not observe this in ASASSN-18tb. 

It is possible the companion interaction models do not accurately represent the early evolution of the \Ha emission. Models in the literature do not provide a clear calculation of when stripped material should start becoming visible. For all spectra with detected \Ha emission, the Fe emission feature at $\lambda\approx 4660$\AA{} is also present, indicating that the inner ejecta are partially visible. However, for the \Ha emission to stem from a stripped companion, the \Ha material would need a previously unincorporated external power source or trapping mechanism to sustain the near-constant luminosity. Furthermore, the early-time \textit{TESS} light curve shows no indication of the excess predicted from ejecta-companion interaction, as discussed in Section~\ref{sec:earlyLC}.

The more likely interpretation appears to be that the H$\alpha$ signature is a product of CSM interaction. An approximately constant H$\alpha$ luminosity prior to $\sim150$ days beyond maximum light is an established feature of SNe~Ia-CSM, and while we do not detect H$\beta$ emission in the spectra we present, the upper limit we place on the Balmer decrement in late phase spectra $(F_{\Ha}/F_{\rm H\beta}\gtrsim 5)$ is consistent with measurements by \cite{2013Silverman} for the SNe~Ia-CSM population. However, even among this rare class of events ASASSN-18tb stands out as a significant outlier in many respects.

A major difference between ASASSN-18tb and other SNe~Ia-CSM is that the light curve of ASASSN-18tb is fairly normal for a low-luminosity SNe~Ia, while other SNe~Ia-CSM generally do not obey the standard empirical SNe~Ia light curve relations \citep{2013Silverman}. \cite{2013Silverman} also found that all SNe~Ia-CSM have absolute magnitudes in the range $-21.3~\rm{mag} \leq M_R \leq -19~\rm{mag}$. ASASSN-18tb is nearly a full magnitude less luminous, with $M_R \approx -18.1~\rm{mag}$. Additionally, while all of the SNe~Ia-CSM identified by \cite{2013Silverman} were found in late-type spirals or dwarf irregulars (star-forming galaxies indicative of young stellar populations), as noted by \cite{2019Kollmeier}, the host of ASASSN-18tb is an early-type galaxy dominated by old stellar populations.

ASASSN-18tb is also spectroscopically distinct from the SNe~Ia-CSM population at early times. While previously identified SNe~Ia-CSM resemble slow-declining, overly luminous 1991T-like SNe~Ia, ASASSN-18tb is more comparable to the fast-declining, underluminous 1999bg-like SNe~Ia. Like SN~1991bg, ASASSN-18tb falls in the ``Cool'' (CL) region of the \cite{2006Branch} Diagram, while SN~1991T and the SNe~Ia-CSM belong to the ``Shallow-Silicon'' (SS) subtype \citep{2019Kollmeier}.

Whether ASASSN-18tb represents a distinct sub-class of SNe~Ia-CSM or the extreme end of a continuum remains to be seen, but it is clearly inconsistent with the properties of previously studied SNe~Ia-CSM. Future observations and theoretical studies of this event will hopefully shed light on its unusual characteristics. X-ray emission has previously been observed for one Ia-CSM SN~2012ca by \cite{2018Bochenek}, and such signatures may be visible for ASASSN-18tb, although the presumably low density CSM of this event would likely make such an observation very challenging. Radio observations are powerful probes of the CSM surrounding SNe \citep{2012Chomiuk,2012Krauss,2013Milisavljevic}, and may be able to better characterize the environment of ASASSN-18tb.

Recent work indicates that underluminous SNe~Ia tend to be produced through the collisional model \citep{2015Dong,2019Vallely}. As shown by \cite{2014Piro}, the combination of the WD mass function and the collisional model simulations of \cite{2013Kushnir} predict a $^{56}$Ni yield distribution peaked near $M_{Ni}\sim0.3\,\text{M}_\odot$, strikingly similar to the $M_{Ni}=0.29\pm0.07~\textnormal{M}_\odot$ we estimate for ASASSN-18tb. As such, it is interesting to ponder a scenario where one might be able to observe CSM interaction from a collisional model DD progenitor scenario. It may be possible to achieve this by invoking a red giant tertiary. 

The collisional model requires a tertiary to drive the eccentricity oscillations that produce the collision. Occasionally the tertiary would be an evolved red giant whose mass loss could produce a low density CSM into which the SN then explodes. While \cite{2013Silverman} found that nearly all SNe~Ia-CSM exhibit H$\alpha$ luminosities in the range $10^{40}-10^{41}$ ergs s$^{-1}$, the \Ha luminosity of \name is 2 orders of magnitude lower at $\sim10^{38}$ ergs s$^{-1}$. This implies an overall lower amount of CSM material for the ejecta to interact with, which can be explained by a tertiary with relatively low mass that has outlived the inner binary. Further observations and theoretical modeling will hopefully constrain the mass of the \Ha emitting material, which can provide additional clues to its origin.

The \textit{TESS} observations we present also emphasize how powerful the mission will be for probing the early-time behavior of SNe. Due to its smaller aperture and wide field of view, \textit{TESS} cannot match \textit{Kepler}'s exquisite precision for events of comparable brightness.
However, \textit{TESS} covers a much larger area of the sky, and will be able to observe significantly more SNe over the duration of its two year mission.
So far, six SNe have been published from the \textit{Kepler} and \textit{K2} missions \citep{2015Olling,2016Garnavich,2019Shappee}, whereas TESS will obtain obtain relatively high precision light curves for $\sim130$ SNe ($\sim$100 SNe~Ia, and $\sim30$ SNe~II; \citealt{2019Fausnaugh}).
These observations will provide an unparalleled sample of early-time SN light curves.

While it is difficult to produce stringent upper limits on companion interaction light curve signatures in a single event (due to the strong viewing angle dependence of the predicted effect), this can easily be accounted for once a large sample of light curves has been collected, and the predicted emission from the \cite{2010Kasen} companion interaction models is readily detectable in the \textit{TESS} band \citep{2019Fausnaugh}. \textit{TESS} will either finally detect the long-sought signature of companion interaction, or put stringent non-detection limits on the phenomenon and add to the growing list of observational constraints in tension with the single-degenerate scenario.

Another advantage of this sample is that because these \textit{TESS} SNe are necessarily bright, it will be possible to obtain late-phase spectra for them. Observations of ASASSN-18bt \citep{2019Shappee,2019Dimitriadis} showed that the early-time light curve alone leads to degeneracies between the observational signatures of the interactions with a nearby companion, radioactive material near the outside of the ejecta, and circumstellar interactions. The combination of a large number of well-observed early-time \textit{TESS} SNe light curves and late-phase spectra of these transients will provide a unique probe that can break these degeneracies.

\section*{Acknowledgments}

We thank the referee, Mark Phillips, for very helpful comments. We thank the Las Cumbres Observatory (LCOGT) and its staff for its continuing support of the ASAS-SN project. This research utilizes LCOGT observations obtained with time allocated through the National Optical Astronomy Observatory TAC  (NOAO Prop. ID 2018B-0110, PI: P. Vallely). We thank Nidia Morrell for obtaining the du Pont spectrum we present. We also thank the Swift PI, the Observation Duty Scientists, and the science planners for promptly approving and executing our observations. Most of the spectroscopic observations were obtained using the Southern African Large Telescope (SALT) in the Rutgers University program 2018-1-MLT-006 (PI: S.~W.~Jha), with an additional SALT spectrum obtained as part of the Large Science Programme on transients (2016-2-LSP-001; PI: Buckley). Polish participation in SALT is funded by grant no. MNiSW DIR/WK/2016/07.

ASAS-SN is supported by the Gordon and Betty Moore Foundation through grant GBMF5490 to the Ohio State University and NSF grant AST-1515927. Development of ASAS-SN has been supported by NSF grant AST-0908816, the Mt. Cuba Astronomical Foundation, the Center for Cosmology and AstroParticle Physics at the Ohio State University, the Chinese Academy of Sciences South America Center for Astronomy (CASSACA), the Villum Foundation, and George Skestos.

PJV is supported by the National Science Foundation Graduate Research Fellowship Program Under Grant No. DGE-1343012. This work at Rutgers University (SWJ, YE) is supported by NSF award AST-1615455. MAT acknowledges support from the DOE CSGF through grant DE-SC0019323. KZS, CSK, and TAT are supported by NSF grants AST-1515876, AST-1515927, and AST-1814440. CSK is also supported by a fellowship from the Radcliffe Institute for Advanced Studies at Harvard University. PC, SD and SB acknowledge Project 11573003 supported by NSFC. Support for JLP is provided in part by FONDECYT through the grant 1191038 and by the Ministry of Economy, Development, and Tourism's Millennium Science Initiative through grant IC120009, awarded to The Millennium Institute of Astrophysics, MAS. MS is supported by a research grant (13261) from the VILLUM FONDEN and a project grant by IRFD (Independent Research Fund Denmark). TAT is supported in part by a Simons Foundation Fellowship and an IBM Einstein Fellowship from the Institute for Advanced Study, Princeton. DAHB's research is supported by the National Research Foundation (NRF) of South Africa. MG is supported by the Polish NCN MAESTRO grant 2014/14/A/ST9/00121. SB is partially supported by China postdoctoral science foundation grant No.2018T110006.

This paper includes data collected by the \textit{TESS} mission, which are publicly available from the Mikulski Archive for Space Telescopes (MAST). Funding for the \textit{TESS} mission is provided by NASA's Science Mission directorate. We thank Ethan Kruse for uploading the TESS FFIs to YouTube, as these videos were invaluable when investigating the systematics in our data. This publication makes use of data products from the Two Micron All Sky Survey, which is a joint project of the University of Massachusetts and the Infrared Processing and Analysis Center/California Institute of Technology, funded by the National Aeronautics and Space Administration and the National Science Foundation. This research uses data obtained through the Telescope Access Program (TAP), which has been funded by the
National Astronomical Observatories of China, the Chinese Academy of Sciences, and the Special Fund for Astronomy from the Ministry of Finance.

This research has made use of the SVO Filter Profile Service (http://svo2.cab.inta-csic.es/theory/fps/) supported from the Spanish MINECO through grant AyA2014-55216. See \cite{2012Rodrigo} for more details on the SVO Filter Profile Service. This work has made use of data from the European Space Agency (ESA) mission {\it Gaia} (\url{https://www.cosmos.esa.int/gaia}), processed by the {\it Gaia} Data Processing and Analysis Consortium (DPAC, \url{https://www.cosmos.esa.int/web/gaia/dpac/consortium}). Funding for the DPAC has been provided by national institutions, in particular the institutions participating in the {\it Gaia} Multilateral Agreement.


\label{lastpage}

\end{document}